\documentclass[%
reprint,
superscriptaddress,
nofootinbib,
 amsmath,amssymb,
 aps,
 showkeys,
]{revtex4-2}

\usepackage{graphicx}
\usepackage{dcolumn}
\usepackage{bm}
\usepackage{hyperref}
\usepackage[bottom]{footmisc}
\usepackage{footnote}
\usepackage{xcolor}
\hypersetup{breaklinks = true, colorlinks = true, citecolor = blue, linkcolor = blue, urlcolor = blue}
\usepackage[shortlabels]{enumitem}
\usepackage{amsmath}

\usepackage[mathlines]{lineno}

\usepackage{xfrac}
\usepackage{booktabs}
\usepackage{multirow}
\usepackage{comment}

\newcommand{\be}{\begin{equation}}
\newcommand{\ee}{\end{equation}}
\newcommand{\bea}{\begin{eqnarray}}
\newcommand{\eea}{\end{eqnarray}}

\setlist[description]{style=nextline, leftmargin=0em, labelindent=0em}

\begin{document}

\title{Measuring spin correlation between quarks during QCD confinement}


\affiliation{Academia Sinica, Nankang, 115, Taipei}
\affiliation{Abilene Christian University, Abilene, Texas   79699}
\affiliation{AGH University of Krakow, FPACS, Cracow 30-059, Poland}
\affiliation{Argonne National Laboratory, Argonne, Illinois 60439}
\affiliation{American University in Cairo, New Cairo 11835, Egypt}
\affiliation{Ball State University, Muncie, Indiana, 47306}
\affiliation{Brookhaven National Laboratory, Upton, New York 11973}
\affiliation{University of Calabria \& INFN-Cosenza, Rende 87036, Italy}
\affiliation{University of California, Berkeley, California 94720}
\affiliation{University of California, Davis, California 95616}
\affiliation{University of California, Los Angeles, California 90095}
\affiliation{University of California, Riverside, California 92521}
\affiliation{Central China Normal University, Wuhan, Hubei 430079 }
\affiliation{University of Illinois at Chicago, Chicago, Illinois 60607}
\affiliation{Chongqing University, Chongqing, 401331}
\affiliation{Creighton University, Omaha, Nebraska 68178}
\affiliation{Czech Technical University in Prague, FNSPE, Prague 115 19, Czech Republic}
\affiliation{Technische Universit\"at Darmstadt, Darmstadt 64289, Germany}
\affiliation{National Institute of Technology Durgapur, Durgapur - 713209, India}
\affiliation{ELTE E\"otv\"os Lor\'and University, Budapest, Hungary H-1117}
\affiliation{Frankfurt Institute for Advanced Studies FIAS, Frankfurt 60438, Germany}
\affiliation{Fudan University, Shanghai, 200433 }
\affiliation{Guangxi Normal University, Guilin, 541004}
\affiliation{University of Heidelberg, Heidelberg 69120, Germany }
\affiliation{University of Houston, Houston, Texas 77204}
\affiliation{Huzhou University, Huzhou, Zhejiang  313000}
\affiliation{Indian Institute of Science Education and Research (IISER), Berhampur 760010 , India}
\affiliation{Indian Institute of Science Education and Research (IISER) Tirupati, Tirupati 517507, India}
\affiliation{Indian Institute Technology, Patna, Bihar 801106, India}
\affiliation{Indiana University, Bloomington, Indiana 47408}
\affiliation{Institute of Modern Physics, Chinese Academy of Sciences, Lanzhou, Gansu 730000 }
\affiliation{University of Jammu, Jammu 180001, India}
\affiliation{Kent State University, Kent, Ohio 44242}
\affiliation{University of Kentucky, Lexington, Kentucky 40506-0055}
\affiliation{Lanzhou University, Lanzhou, 730000}
\affiliation{Lawrence Berkeley National Laboratory, Berkeley, California 94720}
\affiliation{Lehigh University, Bethlehem, Pennsylvania 18015}
\affiliation{Lovely Professional University, Jalandhar - Delhi G.T. Road, Pagwara, Panjab, 144411, India}
\affiliation{Max-Planck-Institut f\"ur Physik, Munich 80805, Germany}
\affiliation{Michigan State University, East Lansing, Michigan 48824}
\affiliation{National Institute of Science Education and Research, HBNI, Jatni 752050, India}
\affiliation{National Cheng Kung University, Tainan 70101 }
\affiliation{Nuclear Physics Institute of the CAS, Rez 250 68, Czech Republic}
\affiliation{The Ohio State University, Columbus, Ohio 43210}
\affiliation{Panjab University, Chandigarh 160014, India}
\affiliation{Purdue University, West Lafayette, Indiana 47907}
\affiliation{Rice University, Houston, Texas 77251}
\affiliation{Rutgers University, Piscataway, New Jersey 08854}
\affiliation{University of Science and Technology of China, Hefei, Anhui 230026}
\affiliation{South China Normal University, Guangzhou, Guangdong 510631}
\affiliation{Sejong University, Seoul, 05006, Korea, Republic Of}
\affiliation{Shandong University, Qingdao, Shandong 266237}
\affiliation{Shanghai Institute of Applied Physics, Chinese Academy of Sciences, Shanghai 201800}
\affiliation{Southern Connecticut State University, New Haven, Connecticut 06515}
\affiliation{State University of New York, Stony Brook, New York 11794}
\affiliation{Instituto de Alta Investigaci\'on, Universidad de Tarapac\'a, Arica 1000000, Chile}
\affiliation{Temple University, Philadelphia, Pennsylvania 19122}
\affiliation{Texas A\&M University, College Station, Texas 77843}
\affiliation{Texas Southern University, Houston, Texas, 77004}
\affiliation{University of Texas, Austin, Texas 78712}
\affiliation{Tsinghua University, Beijing 100084}
\affiliation{University of Tsukuba, Tsukuba, Ibaraki 305-8571, Japan}
\affiliation{University of Chinese Academy of Sciences, Beijing, 101408}
\affiliation{United States Naval Academy, Annapolis, Maryland 21402}
\affiliation{Valparaiso University, Valparaiso, Indiana 46383}
\affiliation{Variable Energy Cyclotron Centre, Kolkata 700064, India}
\affiliation{Warsaw University of Technology, Warsaw 00-661, Poland}
\affiliation{Wayne State University, Detroit, Michigan 48201}
\affiliation{Wuhan University of Science and Technology, Wuhan, Hubei 430065}
\affiliation{Yale University, New Haven, Connecticut 06520}

\author{B.~E.~Aboona}\affiliation{Texas A\&M University, College Station, Texas 77843}
\author{J.~Adam}\affiliation{Czech Technical University in Prague, FNSPE, Prague 115 19, Czech Republic}
\author{L.~Adamczyk}\affiliation{AGH University of Krakow, FPACS, Cracow 30-059, Poland}
\author{I.~Aggarwal}\affiliation{Panjab University, Chandigarh 160014, India}
\author{M.~M.~Aggarwal}\affiliation{Panjab University, Chandigarh 160014, India}
\author{Z.~Ahammed}\affiliation{Variable Energy Cyclotron Centre, Kolkata 700064, India}
\author{A.~K.~Alshammri}\affiliation{Kent State University, Kent, Ohio 44242}
\author{E.~C.~Aschenauer}\affiliation{Brookhaven National Laboratory, Upton, New York 11973}
\author{S.~Aslam}\affiliation{Fudan University, Shanghai, 200433 }
\author{J.~Atchison}\affiliation{Abilene Christian University, Abilene, Texas   79699}
% ...

\collaboration{STAR Collaboration}\noaffiliation

\date{\today}
             
\begin{abstract} 
The vacuum is now understood to possess a rich and complex structure, characterized by fluctuating energy fields~\cite{RevModPhys.61.1} and a condensate of virtual quark-antiquark pairs. The spontaneous breaking of the approximate chiral symmetry~\cite{PhysRev.122.345}, signaled by the nonvanishing quark condensate $\langle q\bar{q}\rangle$, is dynamically generated through topologically nontrivial gauge configurations such as instantons~\cite{RevModPhys.70.323}. The precise mechanism linking the chiral symmetry breaking to the mass generation associated with quark confinement~\cite{Greensite:2011zz} remains a profound open question in Quantum Chromodynamics (QCD) - the fundamental theory of strong interaction. High energy proton-proton collisions could liberate virtual quark-antiquark pairs from the vacuum that subsequently undergo confinement to form hadrons, whose properties could serve as probes into QCD confinement and the quark condensate. Here, we report evidence of spin correlations in $\Lambda\bar{\Lambda}$ hyperon pairs inherited from spin-correlated strange quark-antiquark virtual pairs. Measurements by the STAR experiment at the Relativistic Heavy-Ion Collider (RHIC) at Brookhaven National Laboratory reveal a relative polarization signal of $(18 \pm 4)\%$ that links the virtual spin-correlated quark pairs from the QCD vacuum to their final-state hadron counterparts. Crucially, this correlation vanishes when the hyperon pairs are widely separated in angle, consistent with the decoherence of the quantum system. Our findings provide a new experimental paradigm for exploring the dynamics and interplay of quark confinement and entanglement.

\end{abstract}

\maketitle

In our observable universe, hadrons, such as protons and neutrons, are among the fundamental building blocks of matter that form the physical world. Hadrons are not fundamental particles, but are composite systems made up of quarks and gluons, collectively referred to as partons. One of nature's most profound phenomena is that partons cannot exist as free particles. Instead, the strong force confines partons into hadrons -- a phenomenon known as confinement~\cite{Greensite:2011zz}. 

More rigorously, confinement arises from the disordering of gauge fields at large distances, such that vacuum fluctuations of the chromodynamic fields confine color flux into narrow tubes. Simply put, this mechanism produces a linearly rising potential between static quarks and ensures the absence of colored asymptotic states. This can be essentially illustrated as pulling two quarks apart to a certain distance where, instead of breaking the quarks, more quarks are created.

The Higgs boson, discovered in 2012 at the Large Hadron Collider ~\cite{CMS:2012qbp,ATLAS:2012yve}, was the final missing piece of the Standard Model~\cite{ParticleDataGroup:2024cfk} and helps to explain the origin of mass for fundamental particles such as quarks and leptons. Notably, light quarks by themselves have masses of only a few MeV/${c^2}$, yet protons and neutrons -- each composed of just three valence quarks (up and down) bound by massless gluons -- have masses on the order of $1$ GeV/$c^2$, making them about 150 times more massive. Where does the majority of hadron mass come from? Similarly, the spin structure of the proton presents another puzzle: experimental measurements indicate that quark contributions account for only about 35\% of the total proton spin \cite{EuropeanMuon:1987isl}, in stark contrast to the expectation from the SU(6) quark model~\cite{su6:model}, which predicts that 100\% of the spin arises from valence quarks. The fundamental question is to understand the origin of these emergent hadron structures, e.g. mass and spin, which arise as a consequence of quark confinement.

QCD~\cite{PhysRevLett.30.1343}, the theory of strong interactions, exhibits asymptotic freedom where partons interact weakly in short distances. The absence of such asymptotic states at large distances, e.g., the length scale of a hadron $\sim\!1$ fm ($10^{-15}$ meters), leads to the quark confinement. To understand this, however, the complexity of first-principle QCD calculations becomes difficult to solve numerically at current computational power, due to the nature of low energy self-interacting gluons. This challenging regime, where simple approximations no longer work, is called nonperturbative QCD. Therefore, the detailed mechanisms through which confinement occurs from partons to hadrons, and how it manifests itself in hadron structure, remain unresolved puzzles~\cite{Accardi:2012qut}.

\begin{figure*}[thb]
    \centering
    \includegraphics[width=0.9\textwidth]{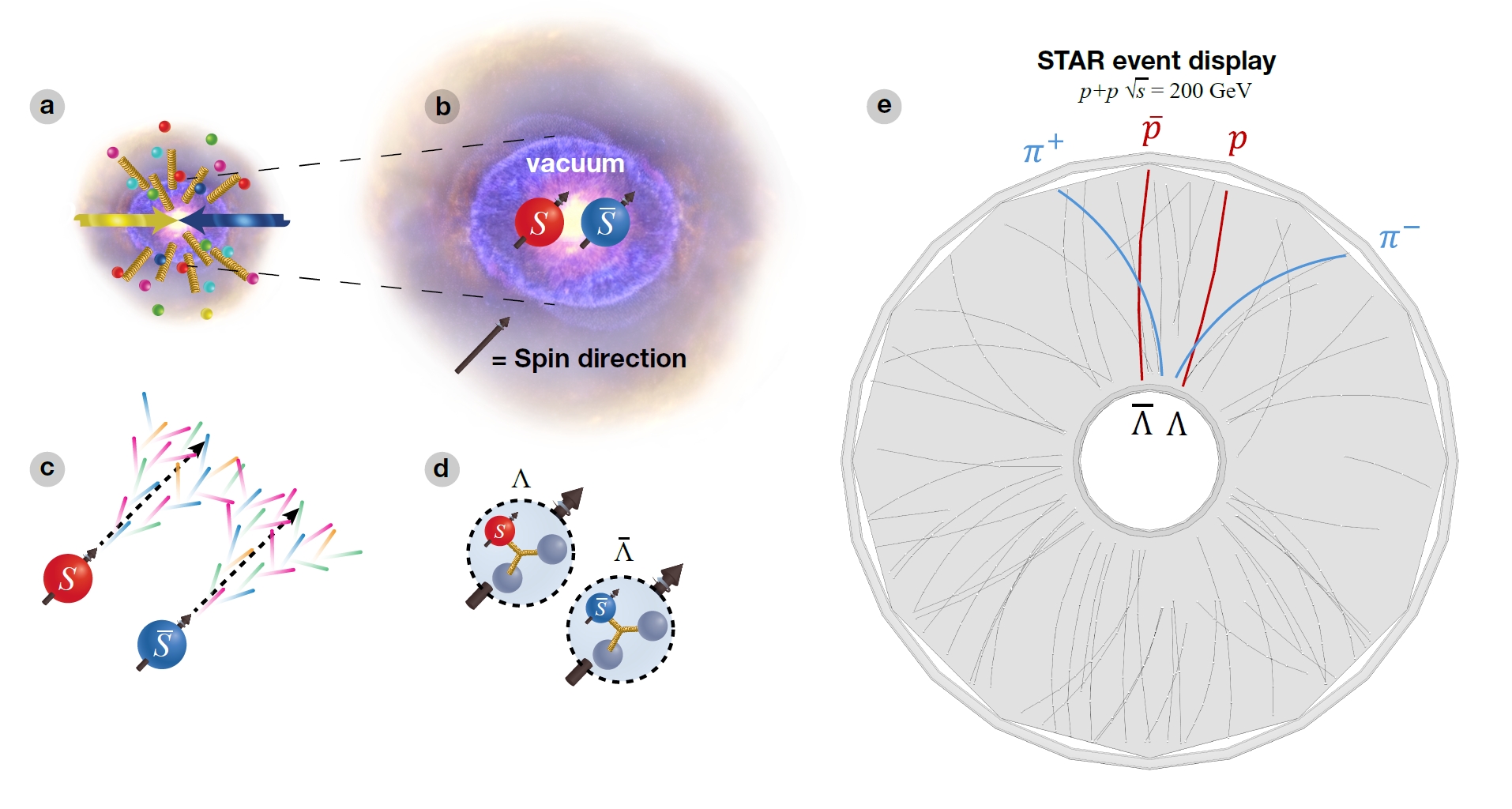}        
    \caption{Illustration of tracing the QCD evolution of the spin of a strange quark-antiquark pair to a $\Lambda\bar{\Lambda}$ hyperon pair and how it can be measured by the STAR experiment at RHIC. See (a)-(e) in text for details.}
    \label{fig_0}
\end{figure*}

\textbf{Analysis Method.} We introduce a novel experimental approach to investigate quark confinement by studying parton evolution and the transition of virtual quarks from the QCD vacuum to final-state hadrons.  Similar to the Higgs mechanism, chiral symmetry is spontaneously broken in the QCD vacuum at zero temperature. It is expected that there are similar numbers of virtual up ($u$), down ($d$), and strange ($s$) quark pairs~\cite{Ellis:1995fc} forming the quark condensate.

Due to the quantum numbers of the vacuum, $J^{PC}=0^{++}$, where $J$, $P$, and $C$ represent total angular momentum, parity, and charge conjugation, respectively, a strong constraint is imposed on the spin configuration of quark-antiquark pairs from the chiral condensate. As a result, these pairs are expected to have their spins parallel~\cite{Ellis:1995fc} in their rest frame, which means that they are in spin triplet states. Thanks to high-energy proton-proton collisions, these virtual quark-antiquark pairs from the vacuum can be liberated and materialise into real particles. Besides other strong experimental indications in hadron spectroscopy~\cite{PhysRev.175.2195,Leutwyler:1996ej} and Lattice QCD calculations~\cite{Hagler:2009ni}, observing the correlated pairs from the vacuum in spin triplet states can be direct experimental evidence of the quark condensate.

Alternatively, depending on the choice of final-state particles, quark-antiquark pairs can also arise from virtual gluon splitting, i.e., $g\rightarrow q\bar{q}$. This process is expected to play a more substantial role in the high energy regime (also known as the perturbative regime)~\cite{Ellis_2012}, offering complementary insights into hadronisation. Nevertheless, understanding the transition from quark pairs to final-state hadrons remains essential for tackling the fundamental problem of quark confinement.

Specifically, our approach is as follows:

\begin{enumerate}[(a)]
    \item Protons are accelerated to 99.996\% of the speed of light for collisions which excite the QCD vacuum~\cite{Peskin:1995ev} and liberate quark pairs from the condensate.
    \item Out of these quark pairs, there are strange quark-antiquark ($s\bar{s}$) pairs with their spins parallel, i.e., in spin triplet states~\cite{Ellis_2012,Ellis:1995fc}. 
    \item Due to confinement the liberated quarks cannot exist independently. Each quark of the $s\bar{s}$ pair will undergo the quark-to-hadron transition known as hadronisation. 
    \item Some $s\bar{s}$ pairs hadronise into $\Lambda$ and $\bar{\Lambda}$ hyperon pairs, where a $\Lambda$ hyperon has one strange ($s$), one up ($u$), and one down ($d$) quark (The structure of the $\bar{\Lambda}$ is similar, but using the antiquarks.). The $\Lambda$ hyperon is a spin-$1/2$ hadron with a lifetime of about $10^{-10}$ seconds, where the spin polarization can be measured via the decay kinematics and direction of the momentum vector of the daughter particles~\cite{PhysRevLett.36.1113}, i.e., proton and pion. From the non-relativistic SU(6) quark model~\cite{su6:model}, the $\Lambda$ hyperon's spin is carried 100\% by the strange quark.
    \item These decay particles, along with other particles, can be measured by the STAR detector. The reconstruction of the decay daughters can provide the spin polarization of the $\Lambda$ and $\bar{\Lambda}$ hyperons, which then allows determination of the hyperon pair spin correlation.
\end{enumerate}
\noindent An illustration of this approach is shown in Fig.~\ref{fig_0}.

This method leverages the spin correlation of $\Lambda\bar{\Lambda}$ hyperon pairs and compares them to their quark-level counterparts. At the moment of $s\bar{s}$ production, the relative spin orientation of the pair is expected to be parallel. During hadronisation, these quarks interact with the surrounding QCD environment to form $\Lambda$ and $\bar{\Lambda}$ hyperons. The novelty of this approach lies in observing the degree of (de)coherence of the correlated $s\bar{s}$ pairs as they transition into hadrons. The quantitative measurement of this (de)coherence provides direct insights into the nonperturbative process of quark-to-hadron transitions, which is challenging for first-principle QCD calculations to address. Tracing this dynamical loss of quantum coherence during hadronisation represents a new paradigm in exploring QCD phenomena.

\textbf{Experiment.} This measurement is performed at the Solenoidal Tracker at RHIC (STAR) detector~\cite{Ackermann:2002ad}. Charged particle tracking, including transverse momentum reconstruction and charge sign determination, is provided by the Time Projection Chamber (TPC) positioned in a 0.5 T solenoidal magnetic field. The TPC volume extends radially from 50 to 200 cm from the beam axis and covers pseudorapidities $|\eta|<1.0$ over the full azimuthal angle, $0<\phi<2\pi$. (Pseudorapidity is a kinematic variable related to the angle ($\theta$) between the particle's momentum and the positive beam axis as, $\eta = -\ln{[\tan{(\theta/2)}}]$. For example, $\eta=1$ corresponds to $\theta \approx 40$ degrees.) The TPC also provides energy loss per unit length ($\mathrm{d}E/\mathrm{d}x$) measurement of tracks used for particle identification.

This measurement was conducted in proton-proton ($p+p$) collisions at the center-of-mass energy $\sqrt{s}=200$~GeV, using a dataset collected in 2012 by the STAR detector at the Relativistic Heavy-Ion Collider (RHIC). All three combinations, $\Lambda\bar{\Lambda}$, $\Lambda\Lambda$, and $\bar{\Lambda}\bar{\Lambda}$, are reported. The data are measured in the two-particle separation in rapidity, $\Delta y$, and azimuthal angle, $\Delta\phi$, respectively. Here, rapidity $y$ is a variable that describes velocity along the beam direction ($y = 1/2 \ln[(E+p_z)/(E-p_z)]$). Data from $K^{0}_{S}K^{0}_{S}$ spin correlations and simulations from the PYTHIA 8.3 Monte Carlo (MC) model \cite{PYTHIA8.3} are compared with the $\Lambda$ measurement and used as a baseline reference. No spin correlation is expected from either of them, as $K^{0}_{S}$ are scalar (spin-0) mesons and no $\Lambda$ hyperon spin physics is included in the PYTHIA 8.3 MC model. The signal extraction method for $K^{0}_{S}K^{0}_{S}$ pairs is the same as for $\Lambda$ hyperon pairs.

\textbf{Data analysis.} Only events with primary vertex within 60~cm from the center of the STAR detector along the proton beam axis were accepted for further analysis. A total of about 600 million minimum-bias $p$+$p$ events were selected and analyzed, requiring the coincidence of STAR Vertex Position Detectors which are located on the upstream and downstream ends of the detector.

    \begin{figure}[ht]
    \centering
    \includegraphics[width=0.5\textwidth]{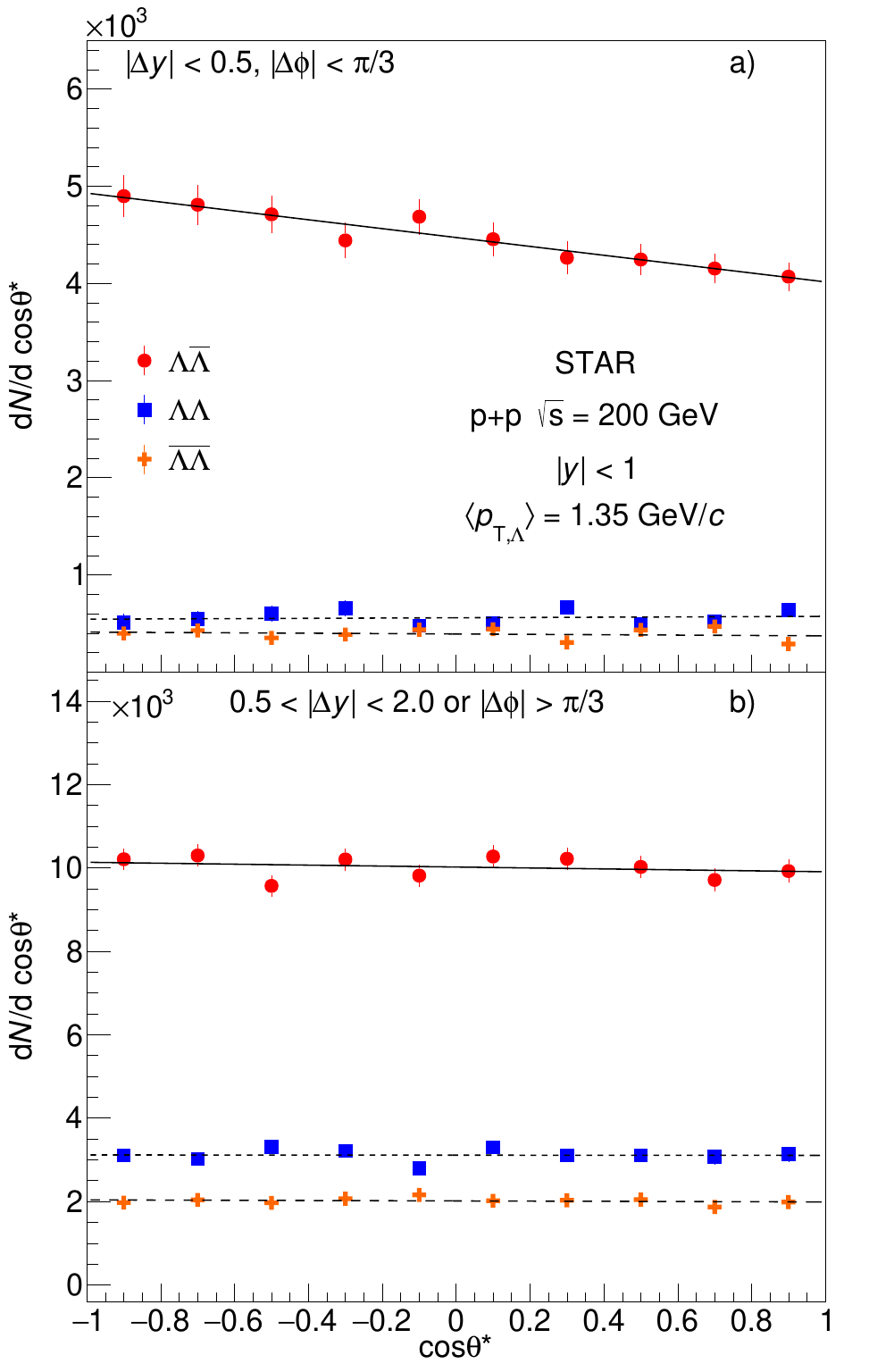}        
    \caption{$\mathrm{d}N/\mathrm{d}\cos\theta^\star$ distributions of decay (anti-)protons for $\Lambda\bar{\Lambda}$, $\Lambda\Lambda$, and $\bar{\Lambda}\bar{\Lambda}$ hyperon pairs measured at mid-rapidity ($|y| < 1$). Panel a) shows the short-range pairs ($|\Delta y| < 0.5$ and $|\Delta \phi| < \pi/3$) and panel b) shows the long-range pairs. Statistical uncertainties are denoted by the error bars. The fits to the data represented by lines are used to demonstrate the magnitude of the spin-spin correlation.}
    \label{fig_2}
\end{figure}

The $\Lambda$ and $\bar{\Lambda}$ hyperons are reconstructed via their hadronic decay $\Lambda \rightarrow p\pi^-$ ($\bar{\Lambda} \rightarrow \bar{p}\pi^+$). The selection of $\Lambda\bar{\Lambda}$, $\Lambda\Lambda$, and $\bar{\Lambda}\bar{\Lambda}$ pairs is done based on a 2-dimensional (2D) Gaussian fit to the 2D invariant mass ($M_\mathrm{inv}$) distributions of the $p\pi$ pairs. Only $\Lambda$ and $\bar{\Lambda}$ hyperon candidates that are at mid-rapidity ($|y|<1$), with transverse momentum $p_\mathrm{T}$ within $0.5<p_\mathrm{T}<5.0\,\mathrm{GeV}/c$ are selected for the analysis. The average transverse momentum $\left <p_\mathrm{T,\Lambda}\right >$ of reconstructed $\Lambda$ hyperons is 1.35 GeV$/c$. 

Based on PYTHIA 8.2~\cite{Sjostrand:2014zea} and STAR detector simulation, only 11\% of measured $\Lambda\bar{\Lambda}$ pairs contain primary $\Lambda$ and $\bar{\Lambda}$ hyperons. The remaining 89\% of the pairs have at least one $\Lambda$ or $\bar{\Lambda}$ hyperon from the decay of a higher mass particle. The impact of this so-called feed-down contribution is included in the model calculations when compared to data (Methods).

    \begin{figure*}[thb]
    \centering
    \includegraphics[width=0.8\textwidth]{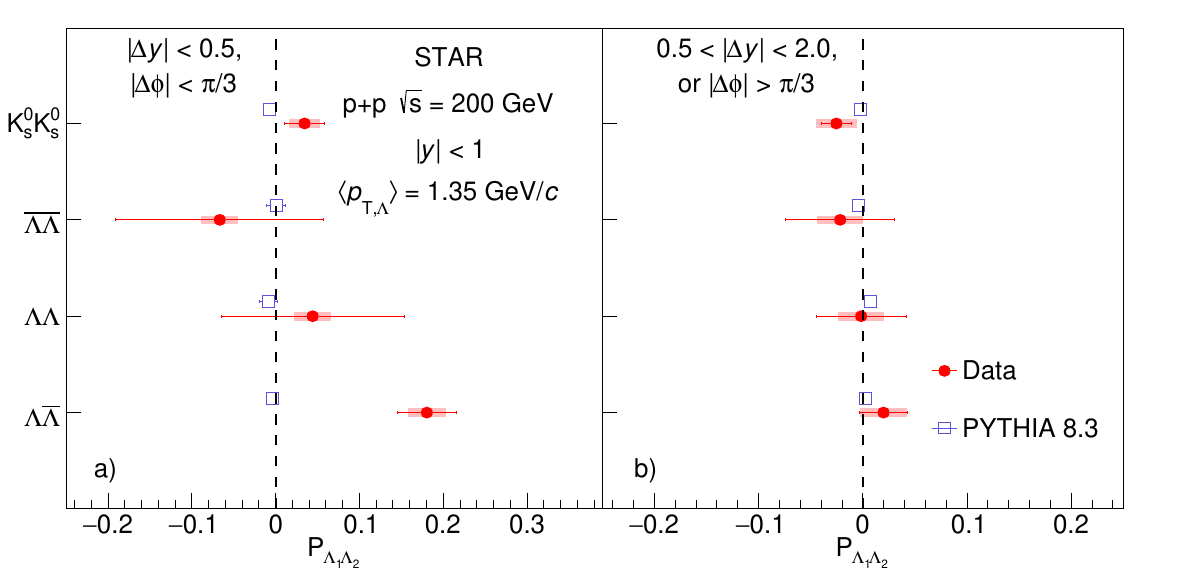}        
    \caption{Spin correlation $P_\mathrm{\Lambda_1\Lambda_2}$ of short-range (left) and long-range (right) $\Lambda\bar{\Lambda}$, $\Lambda\Lambda$, and $\bar{\Lambda}\bar{\Lambda}$ hyperon pairs. The hyperon pair $P_\mathrm{\Lambda_1\Lambda_2}$ is compared to $K^0_\mathrm{S}K^0_\mathrm{S}$ measurements and PYTHIA 8.3 predictions. Statistical uncertainties are denoted by the error bars, and the systematic uncertainties are represented by the shaded boxes.}
    \label{fig_3}
\end{figure*}

\textbf{Measurement of spin correlations.} After selecting the signal $\Lambda$ pairs, the decay (anti-)protons are boosted into the rest frames of their parent particles and the opening angle $\theta^\star$ between the two boosted (anti-)protons is determined. Such (anti-)proton pairs are expected to follow the angular distribution~\cite{Gong:2021bcp,Tornqvist:1980af, Tornqvist:1986pe}:

\begin{equation}\label{eq_1}
    \frac{1}{N}\frac{\mathrm{d}N}{\mathrm{d}\cos\theta^\star} = \frac{1}{2} [1 + \alpha_1\alpha_2P_\mathrm{\Lambda_1\Lambda_2}\cos\theta^\star],
\end{equation}

\noindent where $P_\mathrm{\Lambda_1\Lambda_2}$ is the spin correlation signal or the relative polarization of the $\Lambda$ hyperon pair and $\alpha_1$ and $\alpha_2$ are the weak decay parameters of the $\Lambda$ ($\alpha_- = 0.747\pm0.009$) or $\bar{\Lambda}$ ($\alpha_+ = -0.757\pm0.004$) \cite{ParticleDataGroup:2024cfk}. For parallel spins we expect $P_\mathrm{\Lambda_1\Lambda_2}=1/3$, while for anti-parallel spins $P_\mathrm{\Lambda_1\Lambda_2}=-1$, and for no spin correlation, $P_\mathrm{\Lambda_1\Lambda_2}=0$.

The $\mathrm{d}N/\mathrm{d}\cos\theta^\star$ distribution is constructed for both the total hyperon pairs, which includes signal and background, and the background-only hyperon pairs. Before the signal can be extracted, the raw $\mathrm{d}N/\mathrm{d}\cos\theta^\star$ distributions were corrected for detector acceptance loss and inefficiency. The correction was performed using the mixed-event (ME) technique for all $\mathrm{d}N/\mathrm{d}\cos\theta^\star$ distributions. 

The details of the $\Lambda$ reconstruction, signal extraction, and the ME event corrections are described in the Methods.   

In Fig.~\ref{fig_2}, the corrected $\mathrm{d}N/\mathrm{d}\cos\theta^\star$ signal distributions for $\Lambda\bar{\Lambda}$, $\Lambda\Lambda$, and $\bar{\Lambda}\bar{\Lambda}$ are shown. The top and bottom panels show the spin correlations for short-range ($|\Delta y| < 0.5$ and $|\Delta\phi| < \pi/3$) and long-range ($0.5 < |\Delta y| < 2.0$ and/or $\pi/3 < |\Delta\phi| < \pi$) $\Lambda$ pairs, respectively. The lines are the linear fits to the data according to Eq.~\ref{eq_1}. Quality of the fits used for signal extraction is discussed in Methods.

\textbf{Results.} Figure~\ref{fig_3} shows the $\Lambda$ hyperon spin correlations, expressed in terms of the value of $P_\mathrm{\Lambda_1\Lambda_2}$, for short-range (left) and long-range (right) pairs. It is found that the short-range $\Lambda\bar{\Lambda}$ pairs show a positive spin correlation of $P_\mathrm{\Lambda\bar{\Lambda}} = 0.181 \pm 0.035_\mathrm{stat} \pm 0.022_\mathrm{sys}$, with a 4.4 standard deviation significance with respect to zero. For details about the systematic uncertainties, see Methods. The short-range $\Lambda\Lambda$ and $\bar{\Lambda}\bar{\Lambda}$ pairs and all long-range pairs exhibit spin correlation consistent with zero. The $K^0_\mathrm{S}K^0_\mathrm{S}$ measurements and PYTHIA 8.3 predictions are shown for comparison and are consistent with zero for both short-range and long-range pairs, as expected.  This result marks the first evidence of a positive spin correlation between $\Lambda$ and $\bar{\Lambda}$ in high-energy $p+p$ collisions. 

The positive polarization of short-range $\Lambda\bar{\Lambda}$ pairs corresponds to a parallel spin configuration~\cite{ALEXANDER1996377}, where this orientation of the spin is expected from the chiral condensate $\langle q\bar{q} \rangle \ne0$~\cite{Ellis:1995fc}. An alternative scenario -- gluons splitting into $s\bar{s}$ pairs -- has also been investigated. According to the PYTHIA 8.3 prediction, we found negligible contributions from this process for pairs within our measured momentum range. Furthermore, hadronic final-state interaction has been investigated via a femtoscopic type correlator~\cite{Baym:1997ce}, which is found to be negligible. Therefore, the observed spin correlation is strong evidence for the presence of vacuum quark pairs originating from the chiral condensate.

\begin{figure*}[thb]
    \centering
    \includegraphics[width=0.9\textwidth]{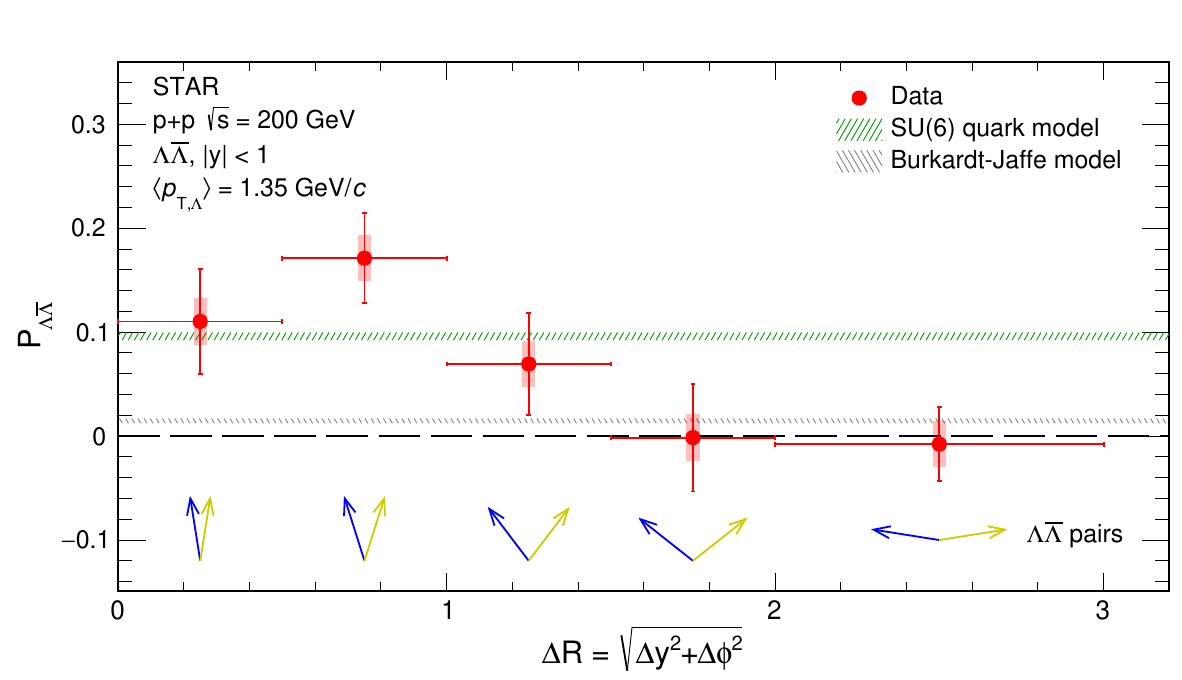}        
    \caption{Spin correlation $P_\mathrm{\Lambda_1\Lambda_2}$ as a function of pair separation $\Delta R$. The data are compared with predictions from the SU(6) quark model~\cite{su6:model} and the Burkardt-Jaffe model~\cite{Burkardt:1993zh}. Statistical uncertainties are denoted by the error bars, and the systematic uncertainties are represented by the shaded boxes. The blue and yellow arrows are used to illustrate the separation of the $\Lambda\bar{\Lambda}$ pairs. }
    \label{fig_4}
\end{figure*}

We have studied the pair kinematic dependence of this spin correlation to further understand the underlying spin correlation. As the separation of the pairs, characterised by $\Delta R=\sqrt{\Delta y^{2}+\Delta\phi^{2}}$, increases the spin correlation of $\Lambda\bar{\Lambda}$ is found to be weaker, as shown in Fig.~\ref{fig_4}. In addition, we compare the results to model calculations in conjunction with the feed-down contributions of $\Lambda$s based on prediction from PYTHIA 8.2 and STAR detector simulation. 

Although the maximum relative polarization correlation of the spin parallel $s\bar{s}$ pair is $P_\mathrm{\Lambda_1\Lambda_2}=1/3$~\cite{Tornqvist:1986pe}, with the feed-down contributions, e.g., from $\Sigma^0$ and other strange baryons, the correlation for $\Lambda\bar{\Lambda}$ pairs is expected to be less than $1/3$. The expected $\Lambda\bar{\Lambda}$ spin correlation from the SU(6) model is found to be $(9.6\pm0.4)\%$ with our data kinematic selections (Methods). The model calculation has no interaction mechanism, so the results only reflect the correlation on the hadron level assuming the initial strange quark pairs are still 100\% spin aligned. We find that the data are compatible with the SU(6) quark model, with 100\% spin aligned $s\bar{s}$ pairs in the initial state, within the uncertainty at small $\Delta R$. The Burkardt-Jaffe model predicts smaller polarization~\cite{Burkardt:1993zh}, and is disfavoured by our data. The detailed calculations of these models, as well as the PYTHIA 8 simulations for feed-down contributions are included in Methods. 

Our result shows a large spin correlation at small $\Delta R$, but a correlation consistent with zero at large $\Delta R$. This suggests that: i) within uncertainties, the spin correlation of short-range $\Lambda\bar{\Lambda}$ pairs are at their maximal values, consistent with inheriting 100\% from their $s\bar{s}$ counterparts at the quark level. ii) Decoherence effects from quark and gluon interactions or multiple initial $s\bar{s}$ pairs might have diluted, if not washed out, the spin correlations when the pairs are widely separated. We expect both findings will contribute to our understanding of QCD evolution and quark-to-hadron transitions.

In terms of entanglement measures, e.g., the Peres–Horodecki criterion or Positive Partial Transpose (PPT) test~\cite{Peres:1996dw,Horodecki:1997vt}, the case of spin–triplet $\Lambda\bar{\Lambda}$ states warrants further detailed investigation. In particular, it is important to evaluate both the isotropic two spin-$\tfrac{1}{2}$ configuration, for which the separability bound can be expressed in terms of a single correlation parameter, and the more general case that requires a full correlation–tensor analysis within the PPT framework. From the experimental side, additional care is also needed to quantify possible feed-down effects that could influence the observed spin correlations.

Similar measurements were performed in the past at experiment PS185 at LEAR, where spin triplet states were observed in the exclusive reaction $\bar{p}p\rightarrow \Lambda\bar{\Lambda}$~\cite{BARNES1993277,PS185:2006yyx}. This fixed-target experiment was conducted with an anti-proton beam at approximately $1.7~\mathrm{GeV}/c$, featuring kinematics significantly different from those in this study. Moreover, the spin correlation was measured via a global axis, i.e., with respect to the production plane. At the current STAR kinematics, global polarization is not expected~\cite{PhysRevD.91.032004}. This was verified by measuring the spin-spin correlation of $\Lambda\bar{\Lambda}$ pairs that are close in $\phi$ and far in $y$ ($|\Delta\phi|<\pi/3$, $|\Delta y|>0.6$). No spin-spin correlation is observed for such pairs, with $P_{\Lambda\bar{\Lambda}} = -0.012 \pm 0.073_\mathrm{stat} \pm 0.022_\mathrm{sys}$, which indicates that the observed spin-spin correlation for short-range $\Lambda\bar{\Lambda}$ pairs is not a result of correlation of $\Lambda$ and $\bar{\Lambda}$ to common global production plane. Establishing the exact connection, however, remains an interesting subject for future investigation in collaboration with theoretical studies. Other noteworthy measurements, similar to ours, are those by BESIII Collaboration which used hyperon spin correlations to look for CP symmetry violation signals in $J/\psi$ decay \cite{BESIII:Nature,BESIII:NaturePhysics}.\\

\noindent \textbf{Discussion and Applications}
\begin{description}
\item[QCD confinement and chiral symmetry breaking] In QCD, confinement and spontaneous chiral symmetry breaking are rigorously defined phenomena~\cite{PhysRev.122.345, Greensite:2011zz}, and our very existence is, in a sense, a testament to their reality. What remains less well understood, however, is their role in the formation of hadrons—the transition from quarks to bound states—and how fundamental properties such as mass and spin emerge in this process.
    
In this work, we present a new experimental approach to study the evolution of spin correlation during the nonperturbative hadronisation process. For the first time, we trace the spin degrees of freedom of a quark–antiquark pair as it evolves into hadrons, demonstrating that most, if not all, the original partonic spin polarization is preserved through hadronisation. By utilizing quark's initial spin correlation, the new experimental approach may provide a more direct probe of the quark condensate to study QCD vacuum structure, e.g., topological charge fluctuations, local strong Charge-conjugate and Parity violation, etc. One immediate implication is to discover if chiral symmetry can be restored (see later chiral symmetry restoration). This finding provides a valuable new probe for lattice QCD calculations and for future quantum computing approaches aimed at unraveling the nonperturbative dynamics of confinement.
    
\item[Spin decomposition] The reported result provides direct experimental insight into how much spin the strange quark can contribute to the $\Lambda$ spin. The result favours the nonrelativistic SU(6) quark model, leaving little room for contributions from gluons and orbital angular momentum. This is counter-intuitive given the famous ``\textit{proton spin crisis}"~\cite{Hansson:2003rf} which suggests that valence quarks only contribute about 35\% of the proton spin. STAR has experimentally confirmed that about half of the remaining 65\% originates from gluons \cite{PhysRevD.105.092011,PhysRevLett.115.092002}. Does the hyperon spin structure exhibit a different decomposition than that of protons? In any scenario, the answer will be important for understanding nonperturbative QCD. 

\item[$\Lambda$ polarization puzzle] One of the outstanding puzzles in nuclear and particle physics is the large transverse $\Lambda$ polarization in unpolarised collisions~\cite{PhysRevLett.36.1113}. Over the past 50 years, the question ``\textit{How does the $\Lambda$ hyperon obtain its spin?}" has been extensively debated. See~\cite{Tu:2023few} for further discussions. The reported result provides a new experimental constraint to validate both initial-state and final-state driven models, especially because the large transverse polarization has been observed in $\Lambda$ production, but not $\bar{\Lambda}$ production \cite{HERMES:2007fpi}.    
   
\item[Spin transfer] This experimental approach provides valuable insights into spin transfer measurements done in the past~\cite{STAR:2018fqv,STAR:2009hex,PhysRevD.109.012004,COMPASS:2009nhs} to measure the quark helicity and transversity distributions (how much of the longitudinal and transverse polarization of the proton propagates to its quarks, respectively). For more information about these distributions, see Ref.~\cite{Boussarie:2023izj}.

\item[Orbital Angular Momentum (OAM)] Due to quantum numbers of the vacuum, the OAM state $L=1$ is expected for the $s\bar{s}$ pairs. From the final state hyperons, one can measure the momentum distribution in the center-of-mass frame of the pair, where the scenarios of $L=0$ and $L=1$ would exhibit different momentum dependence \cite{su6:model}. This is of great interest for future measurements and may link to the problem of quark OAM inside the proton. 
    
\item[Quantum decoherence] We find that the kinematic dependence of the $\Lambda\bar{\Lambda}$ spin correlations may reveal quantum decoherence effects in the $p$+$p$ collision system. We know there is entanglement from top quark pairs~\cite{ATLAS:2023fsd,CMS:2024pts}. In this case, the initial state $s\bar{s}$ pairs could be in a mixed triplet state, which may require a general PPT test that goes beyond just measuring the relative polarization, to perform the entanglement measure. Nonetheless, using pairs measured after hadronisation and going from short-range to long range pairs, the relative reduction in spin correlation may be sensitive to quantum decoherence effect from the initial states. Similarly, according to Ref.~\cite{Gong:2021bcp}, building on Tornqvist’s work~\cite{Tornqvist:1980af,Tornqvist:1986pe}, this could test Bell’s inequality and pair nonlocality to study QCD string spin dynamics. This opens a new paradigm in exploring hadronisation in the context of quantum information science and the quantum-to-classical transitions~\cite{Schlosshauer:2014pgr}. 
    
\item[Chiral symmetry restoration] At high temperatures, QCD matter—such as the quark-gluon plasma (QGP)—is expected to undergo chiral symmetry restoration due to the disappearance of the quark condensate. However, experimental evidence for this phenomenon in heavy-ion collisions remains inconclusive~\cite{CMS:2016wfo,CMS:2017lrw,ALICE:2017sss,STAR:2021mii,Sung:2021myr,piAF:2022gvw,Rapp:1999ej}. The observation of a distinct $\Lambda\bar{\Lambda}$ spin correlation, particularly in spin singlet states (e.g., $^{1}S_0$)~\cite{Ellis_2012}, could serve as a novel experimental probe for investigating QGP dynamics.   
\end{description}

Among future opportunities, one of the most important steps in further understanding the QCD evolution from parton spins to hadron spins is the experimental control of the quark-antiquark spin configuration and its origins. As discussed earlier, the low transverse momentum region is sensitive to the chiral condensate in the QCD vacuum, where $s\bar{s}$ pairs are expected to be 100\% spin aligned. However, as the $\Lambda$ momentum increases, the gluon splitting process, $g\rightarrow s\bar{s}$, becomes more significant. This momentum dependence will be of great interest in the future. Experimental measurements can further constrain these scenarios to higher $\Lambda$ hyperon momentum, to $\Lambda$ as a fragment of a high momentum parton (called jet), or to higher rapidities and/or center-of-mass energies. In addition, momentum correlation function studies~\cite{ALICE:2020mfd,STAR:2014dcy,STAR:2018uho} in conjunction with spin correlations (this work) between $\Lambda$ hyperons (or with other hyperons) can be carried out to explore hadronic final-state interactions, especially in heavy-ion collisions. All of above directions are promising possibilities in the STAR experiment. 

\textbf{Summary.}
We present the first evidence for spin correlations with $\Lambda$ hyperon pairs in high-energy $p+p$ collisions at RHIC, measured across different kinematic regimes. Notably, among all possible combinations of $\Lambda$ hyperon pairs, short-range $\Lambda\bar{\Lambda}$ pairs exhibit a near maximal expected relative polarization, $P_{\Lambda\bar{\Lambda}} = 0.181 \pm 0.035_\mathrm{stat} \pm 0.022_\mathrm{sys}$, with a significance of 4.4 standard deviations. As the pair's separation increases, the spin correlation decreases significantly, likely due to quantum decoherence or other interaction mechanisms. By probing the QCD evolution of a strange quark-antiquark pair that is expected to be spin-aligned from the vacuum condensate, this new hadron-level measurement provides insights into the underlying mechanisms of QCD confinement. The observation of this relative polarization, alongside the methodology established for spin correlation measurements of hyperon pairs, paves the way for a transformative approach to understanding the complex dynamics of QCD.

\clearpage

\begin{center}
    {\textbf{METHODS}}    
\end{center}

\section{Reconstruction of $\Lambda$ hyperons}

For reconstruction of $\Lambda$ hyperons, the first step of the analysis is selection of pure samples of $\pi^+$, $\pi^-$, $p$, and $\bar{p}$. The charged tracks are selected based on their kinematics - transverse momentum $p_\mathrm{T} = \sqrt{p_\mathrm{x}^2 + p_\mathrm{y}^2}$ and pseudorapidity $\eta \equiv - \ln \left[ \tan (\vartheta/2 ) \right] $ - where $\vartheta$ is the angle between the particle momentum and the positive direction of the proton beam ($z$ axis), and their number of hit points inside of the TPC ($N_\mathrm{hits,TPC}$, $N_\mathrm{max,TPC}$). These charged tracks are then identified based on their energy loss in the TPC gas, by limiting the $n\sigma$ variable which quantifies the difference between the measured energy loss and expected energy loss for the hypothesised particle type. The selected proton and pion candidates are then paired, and the pair topology is constrained to identify $\Lambda$ and $\bar{\Lambda}$ hyperon candidates. 

The full selections on $\Lambda$ reconstruction are summarised in Table.~\ref{tab_cuts}. Six topological selection variables are defined as follows: $DCA_\mathrm{p,\pi}$: distance of the closest approach of proton or pion track to the primary vertex (PV). $DCA_\mathrm{pair}$: distance of closest approach of the proton and pion tracks. $DCA_\mathrm{\Lambda}$: distance of closest approach of $\Lambda$ candidate to the PV. $L_\mathrm{dec}$: reconstructed decay length of the hyperon candidate. $\cos\theta$: cosine of the pointing angle $\theta$, where $\theta$ is measured between the reconstructed momentum and the vector connecting the PV to the decay point.

\begin{table}[ht]
    \centering
    \begin{tabular}{cc}
      \toprule
      \multirow{4}{*}{Track selection} & $p_\mathrm{T} > 150\,\mathrm{MeV}/c$ \\
         & $|\eta| < 1$ \\
         & $N_\mathrm{hits,TPC} > 20$\\
         & $N_\mathrm{hits,TPC}/N_\mathrm{max,TPC} > 0.52$\\
      \midrule
      \multirow{2}{*}{Particle identification}  & $|n\sigma_\mathrm{\pi}| < 3$ \\
         & $|n\sigma_\mathrm{p}| < 2$ \\
      \midrule
      \multirow{5}{*}{$\Lambda$ topology} & $DCA_\mathrm{\pi} > 0.3\,\mathrm{cm}$ \\
         & $DCA_\mathrm{p} > 0.1\,\mathrm{cm}$ \\
         & $DCA_\mathrm{pair} < 1.0\,\mathrm{cm}$\\
         & $DCA_\mathrm{\Lambda} < 1.0\,\mathrm{cm}$\\
         & $2\,\mathrm{cm} < L_\mathrm{dec} < 25\,\mathrm{cm}$ \\
         & $\cos\theta > 0.996$  \\
      \bottomrule
    \end{tabular}
    \caption{Selection criteria for $\Lambda$ and $\bar{\Lambda}$ hyperons. Reconstruction relies on track quality, proton and pion identification in the TPC, and the characteristic $\Lambda$ decay topology.}
    \label{tab_cuts}
\end{table}

Lastly, the $K^0_\mathrm{S}$ candidates are reconstructed using a similar topological method. For details, see Ref.~\cite{STAR:2018fqv,STAR:2025jwc}.

\section{$\Lambda$ pairs signal extraction}
To extract the signal of $\Lambda$ candidates, two sets of distributions are filled for each of the $\Lambda$ hyperon pairs.

\begin{figure*}[ht]
    \centering
    \includegraphics[width=\textwidth]{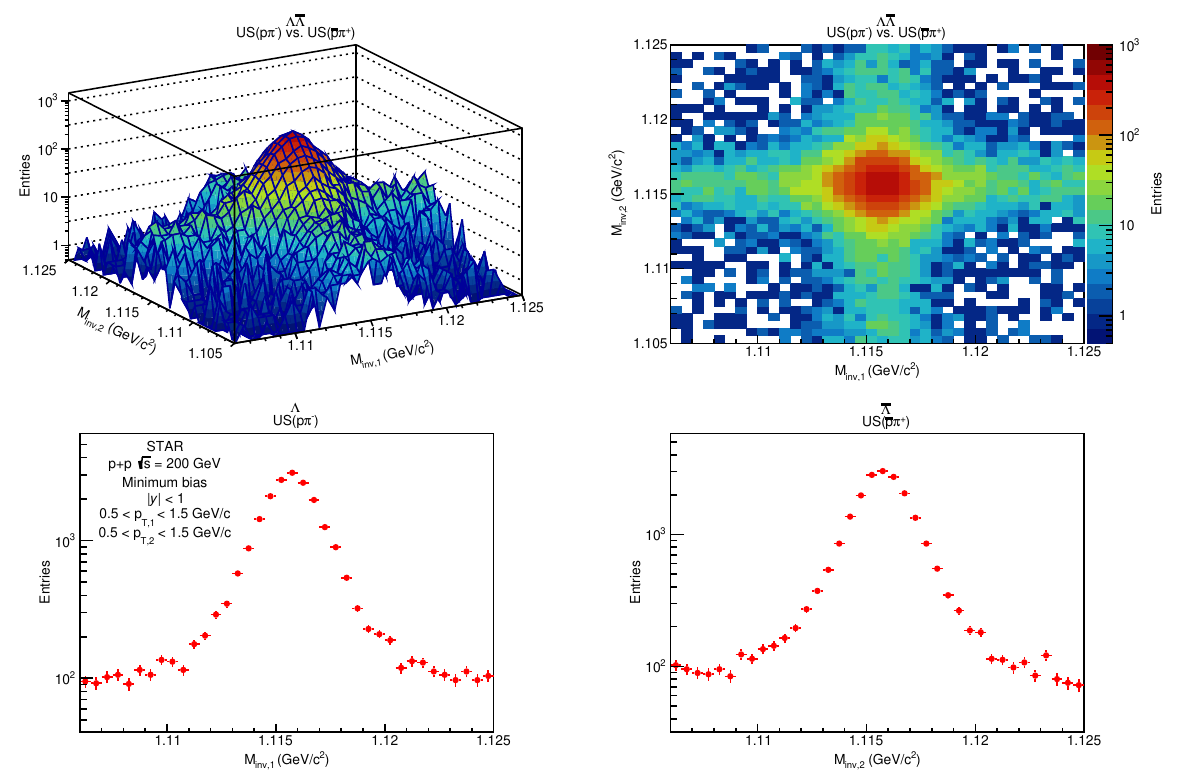}        
    \caption{3D and 2D invariant mass distributions of $p\pi^-$ pairs, paired with $\bar{p}\pi^+$ pairs are shown in the top left and top right panels, respectively. The projections of the multi-dimensional distributions to $p\pi^-$ and $\bar{p}\pi^+$ are shown in the bottom left and bottom right panels, respectively.}
    \label{fig_1}
\end{figure*}

First, an invariant mass, $M_\mathrm{inv}$, distribution that includes an unlike-sign (US) $p\pi$ pair matched with a different US $p\pi$ pair from the same event is obtained. An example of this distribution for $\Lambda\bar{\Lambda}$ pair candidates is shown in Fig.~\ref{fig_1}. The US-US $M_\mathrm{inv}$ distribution has three components: a) the main peak, where two pairs of $p\pi$ decayed from two $\Lambda$ particles from the same event; b) two ridges that correspond to a $p\pi$ pair from a $\Lambda$ decay paired with a combinatorial background pair; c) a continuum that originates from a combinatorial background $p\pi$ pair matched to a different background $p\pi$ pair. 

Second, a $M_\mathrm{inv}$ distribution is constructed by a US $p\pi$ pair and a like-sign (LS) $p\pi$ pair. The US-LS mass distribution is to estimate the two background contributions. It is then subtracted from the US-US distribution, leaving an $M_\mathrm{inv}$ distribution containing only the $\Lambda$ hyperon candidates. The subtracted $M_\mathrm{inv}$ distribution is subsequently fitted with a 2D Gaussian function. Only pairs within $\pm2\sigma$ around the mean are selected for further analysis. The same selection procedure is repeated for $K^0_\mathrm{S}$ mesons. All of the aforementioned distributions are constructed using four distinct particles.

Only $\Lambda$ and $\bar{\Lambda}$ hyperon candidates that are at mid-rapidity ($|y|<1$), with transverse momentum $p_\mathrm{T}$ within $0.5<p_\mathrm{T}<5.0\,\mathrm{GeV}/c$ are selected for the analysis. The average transverse momentum $\left <p_\mathrm{T}\right >$ of the reconstructed $\Lambda$ hyperons is 1.35 GeV$/c$ (for $K^0_\mathrm{S}$ mesons it is 1.14~GeV$/c$). Further in the analysis, the selected pairs are divided into groups based on their relative kinematics ($\Delta \phi$, and $\Delta y$). The numbers of selected signal hyperon pairs is summarized in Tab. \ref{tab_signal}. The signal to background ratios ($S/B$) of the selected hyperon pairs do not heavily depend on this relative kinematics and are in the range of $7<S/B_\mathrm{\Lambda\bar{\Lambda}}<8$ for $\Lambda\bar{\Lambda}$ pairs and in the range of $3<S/B_\mathrm{\Lambda\Lambda}<4$ for both $\Lambda\Lambda$ and $\bar{\Lambda}\bar{\Lambda}$ pairs.

\begin{table}[ht]
    \centering
    \begin{tabular}{cccc}
      \toprule
      Pair kinematics & $N_{\Lambda\bar{\Lambda}}$ & $N_{\Lambda\Lambda}$ & $N_{\bar{\Lambda}\bar{\Lambda}}$ \\
      \midrule
      $|\Delta y|<0.5$, $|\Delta \phi|<\pi/3$ & 8,994 & 1,119 & 786 \\
      $|\Delta y|>0.5$, or $|\Delta \phi|>\pi/3$ & 20,051 & 6,231 & 4,029 \\
      \bottomrule
    \end{tabular}
    \caption{Counts of selected ${\Lambda\bar{\Lambda}}$, ${\Lambda\Lambda}$, and ${\bar{\Lambda}\bar{\Lambda}}$ signal pairs. Both short-range and long-range pairs included in the analysis are listed.}
    \label{tab_signal}    
\end{table}

\section{Mixed event correction}
Before the correlation signal can be extracted, the raw $\mathrm{d}N/\mathrm{d}\cos\theta^\star$ distributions have to be corrected for detector acceptance loss and inefficiency. The dominant detector effect originates from the low momentum cut off on pion $p_\mathrm{T}$. The correction is performed using the mixed-event (ME) technique for all $\mathrm{d}N/\mathrm{d}\cos\theta^\star$ distributions, where the basic assumption is that all detector effects relevant for the $\mathrm{d}N/\mathrm{d}\cos\theta^\star$ shape affect both the same-event (SE) and the ME in the same way. The ME pairs are defined analogously to the SE distributions, except each $\Lambda$ ($\bar{\Lambda}$) in the pair originates from two different events. For example, for one $\Lambda\bar{\Lambda}$ SE pair, we use the $\Lambda$ particle as a reference, and loop over other events that have a $\bar{\Lambda}$ particle. (We also perform the reverse by using $\bar{\Lambda}$ particle as reference.) The ME pairs are selected such that their relative kinematics matches the SE pairs ($|\Delta p_\mathrm{T}| < 0.1\,\mathrm{GeV}/c$, $|\Delta \phi|<0.1$, and $|\Delta y|<0.1$). This is essential in describing the detector effect, because the relative kinematics of the SE hyperon pairs dictates the magnitude of the acceptance effect. In order not to create a bias that some SE pairs will find more ME counterparts than others due to the relative kinematic selections, the ME pairs are weighted by the inverse of the number of times each SE pair is used.

After mixing particles ($\Lambda$ or $K^0_\mathrm{S}$) from different events, the ME distribution of $\mathrm{d}N/\mathrm{d}\cos\theta^\star$ can be used to apply as a correction to the SE distribution. First the ME distributions are normalised to the same number of pairs as the SE. Then the SE distribution is divided by the ME distribution and the resulting, corrected, distribution is re-scaled back to the same statistics as the original, uncorrected, SE distribution.

This acceptance correction method was verified using simulated minimum-bias $p$+$p$ collisions at $\sqrt{s} = 200\,\mathrm{GeV}$ generated by the PYTHIA 8.3 event generator in the default tune. This closure test was performed both for default PYTHIA 8.3, with no expected spin-spin correlation, as well as for PYTHIA 8.3 with an artificially introduced signal. The result was used as an estimate of the systematic uncertainty of the ME technique, discussed in Sec. \ref{sec_sys_err}.

\section{Extracting spin correlation signal}

After correcting the detector effects, the corrected $\mathrm{d}N/\mathrm{d}\cos\theta^\star$ distributions are fitted by Eq.~1 in the main text, and the polarization parameter $P_{\Lambda_{1}\Lambda_{2}}$ is extracted. The quality of the fits was checked by calculating $\chi^2/NDF$ ($NDF$ is number of degrees of freedom of the fit). The resulting fits have all similar $\chi^2/NDF$ values with average of $\chi^2/NDF=0.7$ over all performed fits. The total and background-only $\mathrm{d}N/\mathrm{d}\cos\theta^\star$ distributions provide the spin correlation for a mixture of signal$+$background ($P_\mathrm{S+B}$) and the background only ($P_\mathrm{B}$), respectively. The signal polarization ($P_\mathrm{S}$) is obtained by using the relation:

\begin{equation}\label{eq_signal_P}
    P_\mathrm{S+B} = f_s\cdot P_\mathrm{S} + (1-f_s)\cdot P_\mathrm{B},
\end{equation}

\noindent where $f_s$ and $1-f_s$ are the signal and background fractions, respectively. All background contributions $P_\mathrm{B}$ were found to be consistent with zero. The same analysis procedure is performed for $K^0_\mathrm{S}K^0_\mathrm{S}$ pairs.

\section{Systematic uncertainty}\label{sec_sys_err}
Different sources of systematic uncertainty on the spin correlation of $\Lambda$ hyperon and $K^0_\mathrm{S}$ pairs are considered. The low $p_\mathrm{T}$ cut off on the pion momentum selection is varied from  $p_\mathrm{T} > 150\,\mathrm{MeV}/c$ to $p_\mathrm{T} > 170\,\mathrm{MeV}/c$, which results in an absolute systematic uncertainty of 0.010 in the extracted signal of $P_{\Lambda_{1}\Lambda_{2}}$. The systematic variation of the topological selection of the secondary vertex is modified from the default values $DCA_\mathrm{pair} < 1.0\,\mathrm{cm}$, $2\,\mathrm{cm} < L_\mathrm{dec} < 25\,\mathrm{cm}$, and $\cos\theta > 0.996$ to $DCA_\mathrm{pair} < 0.9\,\mathrm{cm}$, $3\,\mathrm{cm} < L_\mathrm{dec} < 25\,\mathrm{cm}$, and $\cos\theta > 0.997$. This leads to an absolute uncertainty of 0.013 in $P_{\Lambda_{1}\Lambda_{2}}$. Similarly, daughter topological selection was varied from $DCA_\mathrm{p}> 0.1\,\mathrm{cm}$ and $DCA_\mathrm{\pi}> 0.3\,\mathrm{cm}$ to $DCA_\mathrm{p}> 0.2\,\mathrm{cm}$ and $DCA_\mathrm{\pi}> 0.4\,\mathrm{cm}$ which gives an absolute systematic uncertainty of 0.001. The last selection criteria variation was done for $DCA_\mathrm{\Lambda}$ from the analysis value of $DCA_\mathrm{\Lambda}< 1.0\,\mathrm{cm}$ to $DCA_\mathrm{\Lambda}< 0.8\,\mathrm{cm}$ and $DCA_\mathrm{\Lambda} < 1.2\,\mathrm{cm}$ which gives an absolute systematic uncertainty of 0.004. Another systematic uncertainty source comes from the ME correction for the detector effect. This is performed based on a MC simulation using the PYTHIA 8.3 model. The ME-corrected $P_{\Lambda_{1}\Lambda_{2}}$ in PYTHIA 8.3, expected to be zero as there is no genuine spin correlation in the MC model, is checked against the null expectation. A residual polarization value of 0.014 is observed after the ME correction, which is quoted as an absolute uncertainty. Finally, the uncertainties within the quoted values of the weak decay constants are propagated to the final results. The total systematic uncertainty is obtained by adding individual uncertainty sources in quadrature.

\section{Calculations of SU(6) and Burkardt-Jaffe model}
\begin{figure}[ht]
    \centering
    \includegraphics[width=0.5\textwidth]{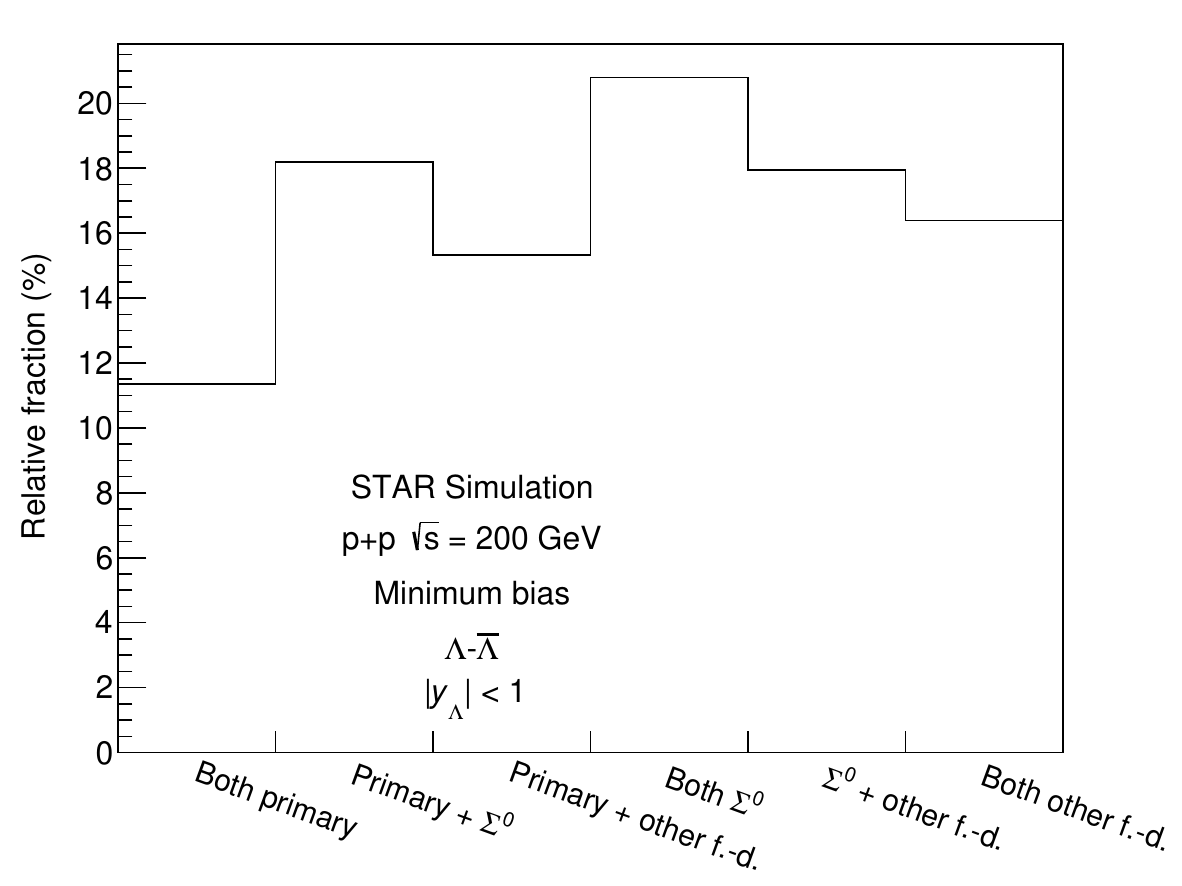}        
    \caption{Relative contribution to $\Lambda\bar{\Lambda}$ hyperon pair feed-down from PYTHIA 8.2 event generator events, which are simulated through the STAR detector. The $\Lambda$ and $\bar{\Lambda}$ hyperons are selected using selection criteria from Tab. \ref{tab_cuts}. The pairs are divided into groups based on the origin of the single $\Lambda$ ($\bar{\Lambda}$) hyperons in the pair: primary $\Lambda$, $\Lambda$ from decay of $\Sigma^0$, and $\Lambda$ from decay of any other hyperon species (labeled as "other f.-d." in axis label).}
    \label{fig_feed_down}
\end{figure}

Based on $p$+$p$ events at $\sqrt{s} = 200\,\mathrm{GeV}$ generated using PYTHIA 8.2, filtered through the STAR detector simulation, we obtain the composition of all $\Lambda$ particles in terms of primary $\Lambda$s and their feed-down contributions. The dominant feed-down contribution is from $\Sigma^0$, which rapidly decays into $\Lambda+\gamma$. With this information, we make a total six of categories of $\Lambda$ hyperon pairs, shown in Fig.~\ref{fig_feed_down}. The plotted percentage is the relative fraction of the total number of $\Lambda\bar{\Lambda}$ pairs that are predicted by the simulation.

We then calculate the maximum expected $\Lambda\bar{\Lambda}$ pair spin-spin correlation using the relative contributions from Fig. \ref{fig_feed_down} and the expected single $\Lambda$ hyperon's polarizations based on the SU(6) and Burkardt-Jaffe (BJ) models~\cite{Burkardt:1993zh}. The single $\Lambda$ hyperon polarization, depending on its parent, is summarized in Table~\ref{tab:su6}. To simplify the calculations, except for $\Sigma^0$, other feed-down contributions are roughly 57\% and -37\% for the SU(6) quark model and the Burkardt-Jaffe model, respectively. The maximum expected pair spin-spin correlation, based on these two models, is then calculated according to the following formula:

\begin{equation}
    P_{\Lambda\bar{\Lambda},\textrm{SU(6)/BJ}} = \frac{1}{3} \cdot \sum_i R_i P_{\Lambda,\textrm{SU(6)/BJ}} P_{\bar{\Lambda},\textrm{SU(6)/BJ}}, 
\end{equation}

\noindent where $R_i$ are the relative feed-down contributions of the $\Lambda\bar{\Lambda}$ pairs and $P_{\Lambda,\textrm{SU(6)/BJ}}$ and $P_{\bar{\Lambda},\textrm{SU(6)/BJ}}$ are the corresponding single $\Lambda$ polarizations from Tab. \ref{tab:su6}. The factor $1/3$ comes from the maximum relative spin polarization given our experimental method~\cite{Tornqvist:1986pe}. We also assign an uncertainty on these values, which comes from the composition of different feed-down contributions in different kinematic regions. In this way we have estimated the maximum expected spin-spin correlation from the SU(6) to be:

\begin{equation}
    P_{\Lambda\bar{\Lambda},\textrm{SU(6)}} = 0.096\pm0.004,
\end{equation}

\noindent and similarly for the BJ model:

\begin{equation}
    P_{\Lambda\bar{\Lambda},\textrm{BJ}} = 0.015\pm0.002.
\end{equation}

\begin{table}[ht]    
    \fontsize{11}{14}\selectfont
    \begin{tabular}{ccc}
    \toprule
    $\Lambda$'s parent             & SU(6) & Burkardt-Jaffe  \\ 
    \midrule
    $s$ quark & 1     & 0.63                \\ 
    $\Sigma^0$                     & 1/9   & 0.15                \\ 
    $\Xi^{0}$                      & 0.6   & -0.37               \\ 
    $\Xi^{-}$                      & 0.6   & -0.37               \\ 
    $\Sigma^{\ast}$                & 5/9   & N/A                  \\ 
    \bottomrule
    \end{tabular}
    \caption{Model predictions for $\Lambda$ hyperon polarization. Shown are the expected $s$ quark contributions from primary $\Lambda$ production and the leading feed-down components in the SU(6) and Burkardt–Jaffe models. Values are taken from Ref.~\cite{Ellis:2007ig}.}
    \label{tab:su6}
\end{table}

In addition, we should note that the PYTHIA 8 MC model generally describes the hyperon productions in proton-proton collisions, while the specific intrinsic uncertainty on the feed-down estimate is unknown and thus not considered here. 

\textbf{Acknowledgement.} We thank the RHIC Operations Group and SDCC at BNL, the NERSC Center at LBNL, and the Open Science Grid consortium for providing resources and support.  This work was supported in part by the Office of Nuclear Physics within the U.S. DOE Office of Science, the U.S. National Science Foundation, National Natural Science Foundation of China, Chinese Academy of Science, the Ministry of Science and Technology of China and the Chinese Ministry of Education, NSTC Taipei, the National Research Foundation of Korea, Czech Science Foundation and Ministry of Education, Youth and Sports of the Czech Republic, Hungarian National Research, Development and Innovation Office, New National Excellency Programme of the Hungarian Ministry of Human Capacities, Department of Atomic Energy and Department of Science and Technology of the Government of India, the National Science Centre and WUT ID-UB of Poland, German Bundesministerium f\"ur Bildung, Wissenschaft, Forschung and Technologie (BMBF), Helmholtz Association, Ministry of Education, Culture, Sports, Science, and Technology (MEXT), and Japan Society for the Promotion of Science (JSPS).

\bibliography{reference_new}

\end{document}